\documentstyle[pre,multicol,aps]{revtex}
\begin{document}
\input{epsf.tex}
\epsfverbosetrue
\title{Vector solitons in (2+1) dimensions}
\author{Johan N. Malmberg, Andreas H. Carlsson, Dan Anderson, and Mietek Lisak}
\address{Department of Electrodynamics, Chalmers University of Technology, S-412 96 G{\"o}teborg, Sweden}
\author{Elena A. Ostrovskaya and Yuri S. Kivshar}
\address{Optical Sciences Centre, The Australian National University, Canberra
ACT 0200, Australia}
\maketitle

\begin{abstract}
We address the problem of existence and stability of vector spatial solitons formed by two incoherently interacting optical beams in bulk Kerr and saturable media. We identify families of (2+1)-dimensional two-mode self-trapped beams, with and without a topological charge, and describe their properties analytically and numerically.
\end{abstract}
\begin{multicols}{2}
\vspace{-2.0cm}
\narrowtext

Recent experimental observations of multidimensional spatial optical solitons in different types of nonlinear materials \cite{exp} call for a systematic analysis of the self-trapping of light in higher dimensions. When two (or more) fields interact nonlinearly, they can form multi-component trapped states, known as {\em vector solitons}. Vector solitons, first theoretically studied in a (1+1)-D models \cite{manakov}, were observed in birefringent fibers and planar waveguides \cite{1d_exp}. Fabricated waveguiding structure localizes such solitons in one of the two directions transverse to the direction of propagation, hence these solitons are effectively one-dimensional. It is only recently, that the theory and experiments on incoherent interaction and truly two-dimensional self-trapping of beams in a bulk (saturable) medium merged \cite{ind_coh}, indicating progress towards the observation and study of different types of (2+1)-D vector solitons and their interactions. 

The practical possibility of such an observation greatly depends upon the soliton stability in the media with realistic, Kerr or saturable nonlinearity. It is known that scalar (one-component), fundamental (2+1)-D solitons are {\em stable} in saturable media \cite{exp}, but they exhibit {\em critical collapse} in Kerr-type media \cite{Rasm}. However, as in the case of (1+1)-D vector solitons \cite{our}, both existence and stability of multi-dimensional vector solitons are nontrivial issues, which have not been systematically adressed so far.

In this Letter, we study (2+1)-D vector solitons in Kerr and saturable media. We analyze two classes of such solitons. First, we consider solitons formed by the coupling of two fundamental modes; such solitons are always bell-shaped. Secondly, we analyze the coupling between the fundamental mode of one field and the first-order mode (i.e. that carrying a topological charge) of the other field. In the latter case the vector solitons may possess a ring structure and are expected to be analogous to the two-hump (1+1)-D vector solitons recently proved to be stable in a saturable medium \cite{our}.

We consider two incoherently interacting beams propagating along the direction $z$ in a bulk, weakly nonlinear optical medium. For a Kerr medium, the problem is described by the normalized, coupled equations for the slowly varying beam envelopes, $E_1$ and $E_2$, 
\begin{equation}
\label{nls}
i \frac{\partial E_{1,2}}{\partial z} + \Delta_{\perp} E_{1,2} + (|E_{1,2}|^2 + \sigma|E_{2,1}|^2)E_{1,2} = 0,
\end{equation}
where $\Delta_{\perp}$ is the transverse Laplacian, and $\sigma$ measures the relative strength of cross- and self-phase modulation effects. Depending on the polarization of the beams, the nature of nonlinearity, and anisotropy of the material, $\sigma$ varies over a wide range. For a Kerr-type material with nonresonant electronic nonlinearity $\sigma \geq 2/3$, whereas for a nonlinearity due to molecular orientation $\sigma \leq 7$ \cite{boyd}.

We look for solutions of Eqs. (\ref{nls}) in the form
\begin{equation}
\label{E}
 E_1=\sqrt{\beta_1}\,u\,e^{i\beta_1z}e^{im_1\varphi}, \qquad E_2=\sqrt{\beta_1}\,v\,e^{i\beta_2z}e^{im_2\varphi},
\end{equation}
where $\beta_1$ and $\beta_2$ are two independent propagation constants, and $m_{1,2} = 0, \pm1$ are topological charges. Measuring the radial coordinate in the units of $\sqrt{\beta_1}$, and introducing the ratio of the propagation constants, $\lambda = \beta_2/\beta_1$, from Eqs. (\ref{E}) we derive a system of stationary equations for the  radially symmetric, normalized envelopes $u$ and $v$: 
\begin{equation}
\label{n nls}
\begin{array}{l}
{\displaystyle \Delta_{\rm r} u - u + (u^2 + \sigma v^2)u = 0,} \\*[9pt] 
{\displaystyle \Delta_{\rm r} v - \frac{m^2_2}{r^2} v - \lambda v + (v^2 + \sigma u^2)v = 0,} 
\end{array}
\end{equation}
where $\Delta_{\rm r}=(1/r)(d/dr)(r d/dr)$, and we assume $m_1=0$. Following the notations introduced in \cite{our}, we describe all vector solitons (\ref{E}) by their {\em ``state vectors''} $|m_1,m_2\rangle$.

\begin{figure}
\setlength{\epsfxsize}{6.6cm}
\centerline{
\epsfbox{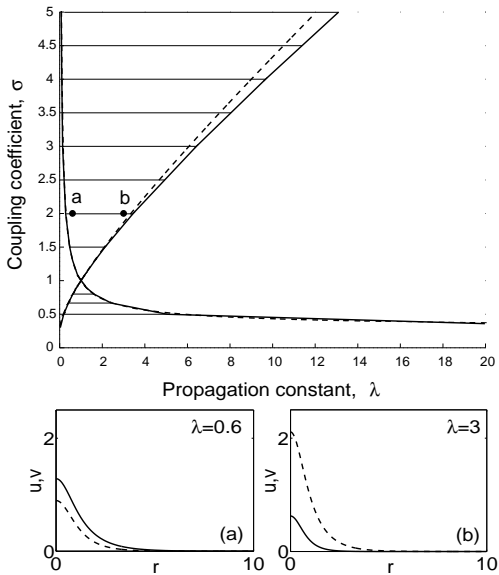}}
\caption{Existence region for the $|0,0\rangle$ vector solitons. Solid - numerically obtained cutoff curves, $\sigma_1(\lambda)$ and $\sigma_2(\lambda)$. Dashed - results of the variational analysis. (a,b) Amplitudes of the $u$ (solid) and $v$ (dashed) components of vector solitons at $\sigma = 2$, corresponding to the points $a$ and $b$ on the plot above.}
\label{fig1}
\end{figure}

First, we consider solutions $|0,0\rangle$. The families of these radially symmetric, two-component vector solitons are characterized by a single parameter $\lambda$, and at any fixed value of $\sigma$, their existence domain is confined between two cut-off values, $\lambda_1$ and $\lambda_2$. When $\lambda < \lambda_1$ or $\lambda > \lambda_2$, self-trapping of coupled fields does not occur, and there exist only scalar solitons for either the $u$ or $v$ components. However, for $\lambda_1 < \lambda < \lambda_2$, a two-mode self-trapped state emerges. Near the cutoff points ($\lambda \approx \lambda_1$ or $\lambda \approx \lambda_2$) this state can be presented as a waveguide created by one field component with a small-amplitude guided mode of the other field component. Examples of $|0,0\rangle$ solitons are presented in Figs. 1(a, b) for $\sigma = 2$. On the parameter plane ($\sigma, \lambda$), the existence domain is skirted by two curves, $\sigma_1(\lambda)$ and $\sigma_2(\lambda)$, which are defined by the corresponding cut-off values, $\lambda_1$ and $\lambda_2$, of the soliton-induced waveguides (see Fig. \ref{fig1}).
For $\sigma = 1$, the $|0,0\rangle$ vector solitons exist only at $\lambda = 1$, and their properties resemble those of the (1+1)-D Manakov vector solitons. They can be constructed by the transformation $u = U\cos\theta$ and $v = U\sin\theta$, where $\theta$ is arbitrary and $U$ satisfies the scalar equation ${d^2U}/{dr^2} + ({1}/{r})(dU/dr) - U +U^3 = 0$.

To describe the existence domain of the multi-dimensional vector solitons analytically, we employ the variational technique \cite{dan}. We look for stationary two-component solutions of Eqs. (\ref{n nls}) in the form $u(r) = A\exp({-r^2/a^2})$, $v(r) = B\exp({-r^2/b^2}),$ where
the parameters $A$, $B$, $a$, and $b$ are defined by variation of the effective Lagrangian of the model (\ref{nls}). Details of this analysis will be published elsewhere. Here, we mention that the coupled algebraic equations of the variational analysis allow us to find the borders of the existence domain for the $|0,0\rangle$ vector solitons: $\sigma_1(\lambda) = (1 + \sqrt{\lambda})^2/4$ and $\sigma_2(\lambda) = (1 + \sqrt{\lambda})^2/(4 \lambda),$
 shown in Fig. \ref{fig1} by dashed curves. One can see that the variational approach provides an excellent alternative to the numerics in identifying the existence domains of the vector solitons.

An important physical characteristic of vector solitons of this type is the total power defined as $P = P_u + P_v = 2 \pi \int^\infty_0\,(u^2 + v^2)rdr$, where the partial powers $P_u$ and $P_v$ are the integrals of motion for the model (\ref{nls}). Figures 2 (a,b) show the total power of the (2+1)-D vector solitons vs $\lambda$ for $\sigma = 2/3$ and $\sigma = 2$, respectively.
\vspace{-2mm}
\begin{figure}
\setlength{\epsfxsize}{7.0cm}
\centerline{
\epsfbox{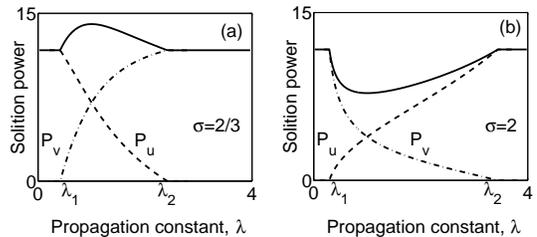}}
\caption{Total power of the vector soliton (solid), and the partial powers of its components (dashed and dot-dashed).}
\label{fig3}
\end{figure}
As follows from these results, the cases $\sigma > 1$ and $\sigma < 1$ are qualitatively different. In the former case, the total power of the vector soliton is lower than the power of a scalar soliton, $P_0 \approx 11.7$ at $\sigma=\lambda=1$ [see Fig. 2(b)]. This is an important and unexpected physical result indicating, in contrast to the commonly held belief, that the excitation of vector solitary waves would require lower input power in comparison to scalar solitons. For $\sigma < 1$, the situation is opposite, and the vector solitons exist at higher power than scalar solitons [see Fig. 2(a)]. A lower total power of the vector soliton, as compared to the corresponding power of a scalar soliton, allows us to explain an effective suppression of the blow-up instability numerically observed at $\sigma=2$, for the special case of $\lambda=1$, when an analytical form of the vector soliton can be found by a Hartree-type ansatz\cite{hayata} .

Similar to the well studied case of the (1+1)-D vector solitons (see, e.g., Ref. \cite{our}), the soliton-induced waveguide can guide higher-order modes. In higher dimensions, such higher-order modes carry a topological charge, i.e. $m_2 \neq 0$ in Eq. (\ref{E}). In the simplest case, we consider $m_2 = \pm 1$ and analyze the system of stationary equations for the radially symmetric wave envelopes, Eqs. (\ref{n nls}). Examples of two-mode $|0,1\rangle$ solitary waves are shown in Figs. 3 (a,b). 
\begin{figure}
\setlength{\epsfxsize}{6cm}
\centerline{
\epsfbox{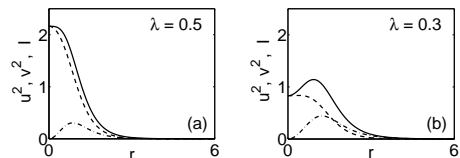}}
\caption{Intensity profiles of vector solitons (solid), formed by scalar components with ($v$, dash-dotted) and without ($u$, dashed) a topological charge ($\sigma = 3$).}
\label{fig4}
\end{figure}
Near the cut-off of the first-order mode, the second component ($v$) appears as a guided mode (dash-dotted) of the effective waveguide created by the scalar soliton in the $u$-component (dashed). The total intensity, shown in Figs. 3 (a) by a solid curve, has a maximum at the beam center. Akin to the humps in the total intensity profile of (1+1)-D solitons \cite{our}, the ring shape of (2+1)-D vector solitons develops far from the cut-off for the first-order mode, when the guided mode deforms the soliton waveguide and creates a coupled state that has the maximum shifted from the beam center [see Fig. 3(b)]. 

Next, we conduct the numerical stability analysis, and an accurate check of the Vakhitov-Kolokolov stability criterion for vector solitons \cite{our}. Our analysis reveals that neither the fundamental nor the first-order mode of the soliton-induced waveguide can arrest the collapse of the scalar bright (2+1)-D solitons in a Kerr medium, and the corresponding vector solitons $|0,0\rangle$ and $|0,1\rangle$ are {\em linearly unstable}. 
\begin{figure}
\setlength{\epsfxsize}{8cm}
\centerline{
\epsfbox{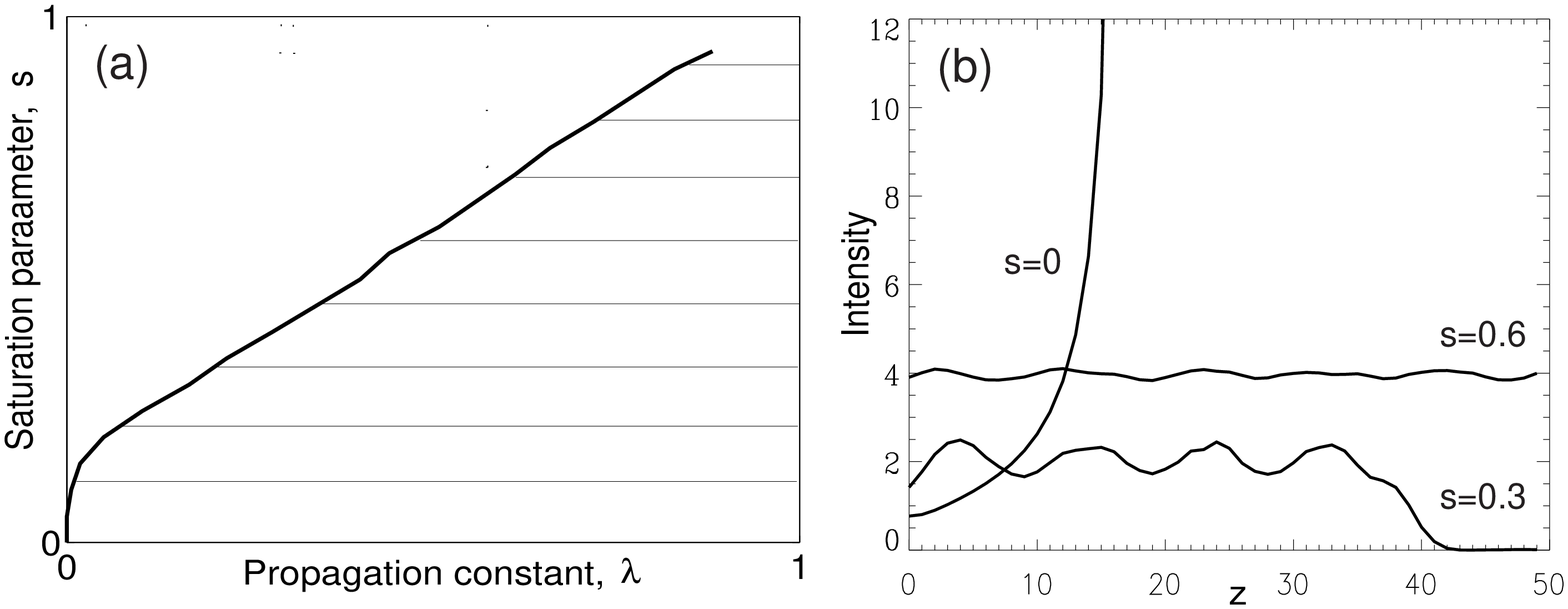}}
\caption{(a) Existence domain for $|0,1\rangle$ solitons in a saturable medium. (b) Evolution of the total intensity at $r=0$ for different values of $s$ ($\lambda=0.6$ and $z=50$).}
\label{fig5}
\end{figure}
\begin{figure}
\setlength{\epsfxsize}{8cm}
\centerline{
\epsfbox{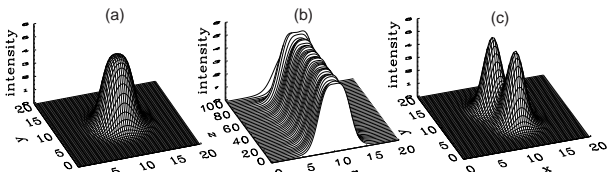}}
\caption{Typical evolution of the $|0,1\rangle$ solitons in a saturable medium ($s=0.65$, $\lambda=0.6$). Shown are (a) the intensity distribution at $z=0$, (b) evolution of the intensity profile at $y=10$, and (c) the intensity distribution at $z=100$. }
\label{fig6}
\end{figure}
Given the established stability of (1+1)-D vector solitons in a {\em saturable medium}\cite{our}, a tempting task is to look for the $|0,1\rangle$ ring-like solitons in such a medium. For the (completely solvable) model of the so-called threshold-type nonlinearity, $|0,1\rangle$ vector solitons have been recently analyzed in \cite{muss}. Here we consider the model corresponding, in the isotropic approximation, to the physically realised solitons in photorefractive materials. The normalized dynamical equations for the envelopes of two incoherently interacting beams can, in this case, be approximately written in the form: $
i \frac{\partial E_{1,2}}{\partial z} + \Delta_{\perp} E_{1,2} + E_{1,2}\left( 1+ |E_{1,2}|^2 + |E_{2,1}|^2 \right) ^{-1} = 0.$
Seeking stationary solutions in the general form (\ref{E}), and introducing the relative propagation constant $\lambda=(1-\beta_2)/(1-\beta_1)$, we arrive, after corresponding renormalizations \cite{our}, to the following system of equations [cf. Eqs. (\ref{n nls})]:
\begin{eqnarray}
\label{sat nls}
\Delta_{\rm r} u - u + uf(I)= 0, \\ \nonumber
\Delta_{\rm r} v - \frac{m^2_2}{r^2} v -\lambda v +vf(I) = 0, 
\end{eqnarray}
where $f(I)=I(1+s I)^{-1}$, $I=u^2+v^2$, and $s=1-\beta_1$ plays the role of a saturation parameter. For $s=0$, this system describes the Kerr nonlinearity (with $\sigma=1$). Since the contributions from the self- and cross-phase modulation are, in this case, equal, the lowest-order bell-shaped $|0,0\rangle$ solutions only exist at $\lambda=1$. In the remaining region of the parameter plane $(s,\lambda)$, the solutions $|0,1\rangle$, similar to those described above for the Kerr nonlinearity, are found [see Fig. 4(a)]. Again, the ring-shaped structure of these solutions develops only far from the cut-off for the vortex-type guided mode. Close to the cutoff, all vector solitons are bell-shaped. 

Although our numerical simulations confirmed that the saturation does have a strong {\em stabilizing} effect on the $|0,1\rangle$  vector solitons [cf. cases $s=0$ and $s=0.6$ in Fig. 4(b)], vector solitons of this type {\em appear to be linearly unstable}. The instability, although largely suppressed by saturation, triggers the decay of the solitons into a dipole structure (as shown in Fig. 5) for even small contribution of the charged mode. However, the quasi-stable dynamics exhibited by these vector solitons over long propagation distances may have serious implications on the attempts to observe them experimentally. Indeed, since, for the current experiments on spatial solitons in photorefractive media \cite{moti}, the propagation length $z=100$ corresponds to a crystal length of up to $40$ mm, this slow-developing dynamical instability may be hard to definitively detect in experiments. 

In conclusion, we have analyzed multi-dimensional vector solitons composed of two incoherently coupled beams in both Kerr and saturable media and defined, analytically and numerically, existence domains for new classes of spatial solitary waves. A stabilizing effect of nonlinearity saturation on (2+1)-D vector solitons with topological charge has been demonstrated.

Yu. S. K. thanks M. Segev for useful discussions and a copy of Ref. \cite{muss} prior to its publication.

\vspace{-2mm}
\end{multicols}
\end{document}